\documentclass{article}
\usepackage[preprint]{nips_2018}
\usepackage[utf8]{inputenc}
\usepackage[T1]{fontenc}
\usepackage{hyperref}   
\usepackage{url}        
\usepackage{booktabs}   
\usepackage{amsfonts}   
\usepackage{nicefrac}   
\usepackage{microtype}
\usepackage{graphicx,subfig}
\usepackage{wrapfig}
\usepackage[numbers]{natbib}

\usepackage{xcolor}

\title{Multi-Domain Processing via Hybrid Denoising Networks for Speech Enhancement}

\author{
  Jang-Hyun Kim$^{*1,2}$, Jaejun Yoo\thanks{Equal contribution.} $^{\ 2}$, Sanghyuk Chun$^{2}$, Adrian Kim$^{2}$, Jung-Woo Ha$^{2}$  \\
  $^{1}$Department of Mathematical Science, Seoul National University\\
  $^{2}$Clova AI Research, NAVER Corp.\\
  \texttt{$^{1}$blue378@snu.ac.kr} \\ \texttt{$^{2}$\{jaejun.yoo, sanghyuk.c, adrian.kim, jungwoo.ha\}@navercorp.com}
}
\begin{document}
\maketitle

\begin{abstract}
We present a hybrid framework that leverages the trade-off between temporal and frequency precision in audio representations to improve the performance of speech enhancement task.
We first show that conventional approaches using specific representations such as raw-audio and spectrograms are each effective at targeting different types of noise.
By integrating both approaches, our model can learn multi-scale and multi-domain features, effectively removing noise existing on different regions on the time-frequency space in a complementary way.
Experimental results show that the proposed hybrid model yields better performance and robustness than using each model individually.
\end{abstract}

\section{Introduction}
The trade-off between temporal and frequency resolution is a well-known phenomenon in the signal processing community,
e.g., the window size in discrete Fourier transformation (DFT) \cite{Hlawatsch1992LinearAQ}.
The larger the time segment, the more frequencies are extracted, thus giving us higher frequency resolution in the expense of temporal resolution. 
Therefore, it is obvious that time-series and time-frequency representations can provide complementary views when investigating a given signal. 
To the best of our knowledge, however,
existing deep learning-based approaches proposed for speech enhancement have only taken either time-series (i.e., raw-audio) \cite{Pascual2017segan, rethage2017wavenet} or time-frequency representation (i.e., spectrogram) as an input \cite{Karjol2018SpeechEU, soni2018time, Zhang2015MultiresolutionSF}.
In this work, we find that models using different audio representations each specialize at tackling specific types of noise, and are also complementary to each other.
Grounding on this observation, we propose a hybrid framework which enables the model to learn multi-scale and multi-domain features, dubbed multi-domain processing via hybrid denoising networks (MDPhD).
We devise a sequential model integrating two modules of both representations by employing auxiliary loss.
Experimental results and ablation studies show that the proposed model can effectively utilize complementary information of time and time-frequency domains. Although our hybridizing strategy is rather straightforward, MDPhD shows better denoising performance than other state of the art (SOTA) algorithms across a variety of noises under multiple measures. Note that the hybrid framework is general and not restricted to the current specific model. The performance can be further improved by employing newly developed models from each domain, by equipping a new loss function, or by designing a better hybridizing strategy.

Our contributions are as follows: 1) We empirically show that the way a model performs denoising depends on its input representation. 2) We propose a hybrid framework that can exploit multi-scale and multi-domain features. To the best of our knowledge, this is the first hybrid approach, effectively utilizing both time and time-frequency domain information. 3) The proposed hybrid model (MDPhD) outperforms SOTA algorithms in the speech enhancement task.

\section{Model Description}

We first describe the objective function and the selected modules that have been reported to show competitive performance using either raw-audio \cite{luo2018tasnet} or spectrogram input \cite{Jansson2017SingingVS}. Selected models are each used later as components of our proposed hybrid model.

\subsection{Objective function}
We employ the energy-conserving loss function proposed in \cite{rethage2017wavenet} which simultaneously considers speech and noise signals. Let the noisy input $x$ consist of clean speech $s$ and noise $n$. 
The estimated speech by the model is referred to as $\hat{s}$. Then, our objective function is defined as follows:
\begin{equation}
    % L(x,s,n,\hat{s}) = |s-\hat{s}| + |n-\hat{n}|,
    L(x,s,n,\hat{s}) = \| s-\hat{s} \|_1 + \| n-\hat{n} \|_1,
\end{equation}
where $\hat{n} = x - \hat{s}$ represents the estimated noise signal and $\|\cdot\|_1$ denotes $\ell_1$ norm.

\subsection{Hybrid Model}
We construct the time domain network based on TasNet \cite{luo2018tasnet} which employs one-dimensional dilated convolution to handle long time sequences of raw-audio. TasNet has shown competitive sample quality for speech source separation, which is a similar task to speech enhancement. In our experiments, we used a reduced version of TasNet. With a slight abuse of notation, we refer to the network as "TasNet" for simplicity. 
For the time-frequency (T-F) domain network, we employ a U-Net structure based on two-dimensional convolutions which has been widely used in various source separation tasks \cite{Jansson2017SingingVS,ronneberger2015u}. The T-F domain network aims to learn an ideal ratio mask (IRM) of a noisy spectrogram input \cite{Wang2014OnTT}. By multiplying the estimated mask to the noisy spectrogram, the model can remove the noise from the time-frequency space. 

\begin{figure}[htbp]
\centering
{\includegraphics[width=0.99\linewidth]{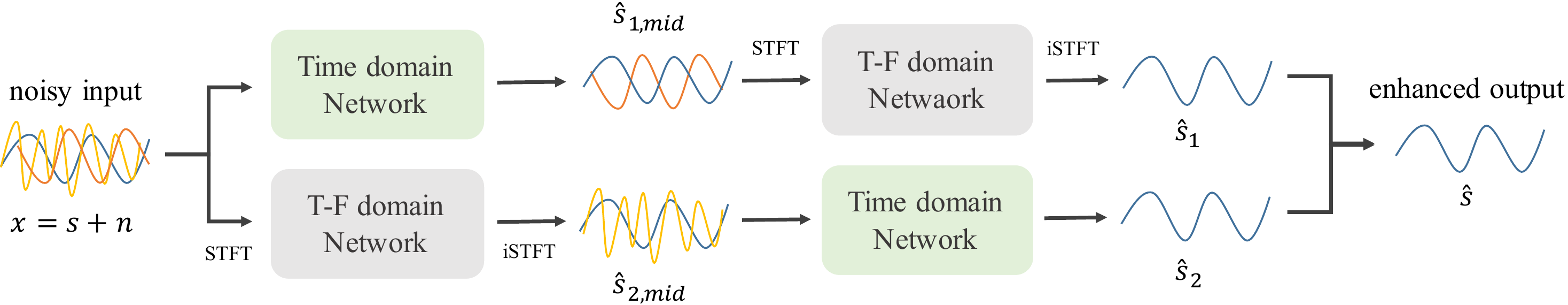}} \\
\caption{A schematic illustration of the hybrid system (MDPhD). Note that the network of the same domain  (same color) shares the parameters. For the time-frequency (T-F) domain network, we convert the time-domain input to a spectrogram using the short time Fourier transform (STFT), whose output is converted back to a waveform using the inverse short time Fourier transform (iSTFT).
}
\label{fig:model20more}
\end{figure}

We hybridize both time and T-F domain networks in a cascaded way (Fig. \ref{fig:model20more}). To make both networks contribute to the denoising task equally well, we devise our model with an auxiliary loss $L(x,s,n,\hat{s}_{i,mid})$ at the intermediate conjunction, where $\hat{s}_{i,mid}$ is the output of the former network. In addition, to let both networks have access to the full data information that is not processed (denoised) by the other, we train the entire model by alternately switching the sequential order of each component. For inference, we can either use a single path or average the results from both paths. Here, we simply average the output of the model, which showed the best performance. 

The final objective of the hybrid model with auxiliary loss becomes
\begin{equation}
    \min_{\theta} \sum_{i=1,2} L(x,s,n,\hat{s}_{i,mid}) + \sum_{i=1,2} L(x,s,n,\hat{s}_{i}),
\end{equation}
where $\theta$ denotes the network parameter.

\section{Experiments}

\subsection{Data and Experimental Setup}
\paragraph{Dataset} We used the dataset \cite{ValentiniBotinhao2016InvestigatingRS} that has been used in the recent speech enhancement studies \cite{Pascual2017segan, rethage2017wavenet}. The dataset was produced by synthesizing the clean speech of Voice Bank corpus \cite{veaux2013voice} and the noise data of Diverse Environments Multichannel Acoustic Noise Database (DEMAND) \cite{thiemann2013diverse}. The training dataset consists of audio from 28 speakers, and the test dataset is composed of the recordings from two speakers. Each speaker's data contains 400 sentences with four noise levels. 
To deal with signals without voices, we added noise-only data to the training dataset, which is a quarter of the total number. 
In our experiments, all audio samples recorded at 48kHz were subsampled to 16kHz.
\paragraph{Experimental Setup} During training and testing, we split speech waveforms with a sliding window of approximately one second (16384 samples) every 500 ms (50\% overlap). 
To obtain the spectrograms, we used the short time Fourier transform (STFT) of 512 window size and 256 hop size with Hanning window. The output spectrogram is converted back to the time domain using the inverse STFT. 
For training, we used batch renormalization to cope with a small batch size of 16 and Adam optimizer with the initial learning rate of 2e-4. The
learning rate was decayed by half every 100,000 iterations. All of the experiments were performed on NAVER Smart Machine Learning (NSML) platform \cite{NSML}.
For more details, please refer to the supplementary material.

\subsection{Experimental Results}

\begin{figure}[b]
\centering
{\includegraphics[width=0.99\linewidth]{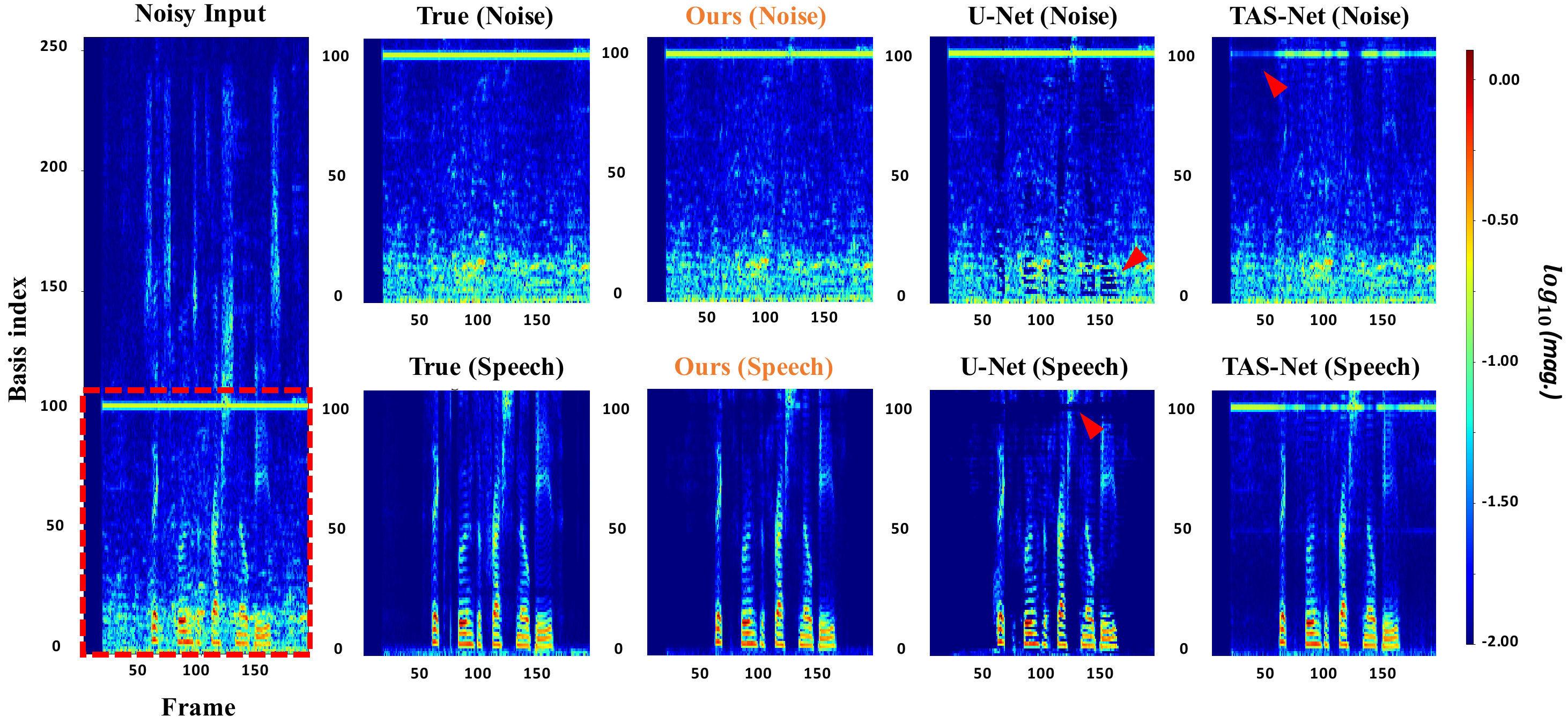}} \\
\caption{Comparison of denoised results for inputs with babble and high frequency noise. For clarity, the output results of the boxed region (red dotted line) of the noisy input is demonstrated in two perspectives. The top row shows the estimated noise and the bottom row displays the estimated speech signal. Some noticeable distortions of U-Net and TasNet in the spectrogram are marked by red arrows.}
\label{figure:noise-b}
\end{figure}

\paragraph{Hybrid Framework Validation} 
To show the complementary characteristic of the time and T-F domain networks, we additionally synthesized noisy signals consisting of speech signals from the test dataset and noises which are either babbles (DEMAND), high frequency sinusoidal noise of 1000 $\sim$ 5000 Hz, or both. Note that the networks did not see any of these noises during the training phase. 
As shown in figure \ref{figure:noise-b}, while the spectrogram approach (U-Net) successfully removes high frequency noise that is prominent in the spectrogram, it suffers from dealing with babble noise which is hardly distinguishable from the frequency components of speech signals.
On the other hand, the raw-audio approach (TasNet) shows superior results on denoising babble noise, which were even better than that of U-Net with doubled parameter size (Table \ref{table:main}). Note that, however, TasNet fails to remove high frequency noise, which is supposedly hard to capture in the time domain (Fig. \ref{figure:noise-b} red arrow).

Table \ref{table:main} summarizes these observations along with ablation studies. Our hybrid model (MDPhD) showed the best performance by combining the strength of each model. 
While the other methods had a noticeable weak domain, MDPhD showed comparable performance across all noise types.
Note that MDPhD showed the best performance when the noises are mixed, which is more practical in real world applications. When we only trained a single path of MDPhD, the model failed to fully utilize the complementary information from both domains. Interestingly, we found that the performance of the model tends to follow the characteristics of the network that comes first in order. For example, the $U\rightarrow D$ model shares the weakness of U-Net and vice versa. We conjecture that this happens because the latter network cannot reconstruct the information that is already lost from the former network. In addition, we tested various objective functions and confirmed that the complementary nature of the two approaches does not come from a specific choice of the objective function (see the supplementary material).

\begin{table}[bt]
\caption{Ablation study performed on various types of noises (babble, high frequency and a mixture of both) with two signal-to-noise ratios (SNRs) (5 and 10 dB). We evaluated the SNR of each model output in decibel (dB) scale. $D$ and $U$ denotes TasNet using one-dimensional dilated convolution and U-Net, respectively. The number of parameters is noted next to the model (e.g., 1.5 M = 1.5 million). $U\rightarrow D$ and $D\rightarrow U$ represent single path models without alternately switching training procedure. Our hybrid model is referred to as $H$ (1.5 M + 1.5 M), where the model exploits both pathways ($U\rightarrow D$ and $D\rightarrow U$) during the training and testing. The best result for each noise type is given in bold style.
}
\renewcommand{\tabcolsep}{0.55mm}
\begin{center}
\begin{tabular}{lcccccc}
\toprule
\multicolumn{1}{c}{} &\multicolumn{2}{c}{\bf babble} &\multicolumn{2}{c}{\bf high freq.} &\multicolumn{2}{c}{\bf babble + high freq.}
\\ \midrule
\multicolumn{1}{c}{} &\multicolumn{1}{c}{\ \ SNR 5} &\multicolumn{1}{c}{\ SNR 10} &\multicolumn{1}{c}{\ \ SNR 5} &\multicolumn{1}{c}{\ SNR 10} &\multicolumn{1}{c}{\ \ SNR 5} &\multicolumn{1}{c}{\ SNR 10}
\\ \midrule
$D$ (1.5 M)& 13.69 & 16.83 & 4.86 & 11.21 & 11.47 & 15.23\\
$D^*$ (3 M)& \textbf{14.25} & \textbf{17.12} & 6.27 & 11.88 & 12.74 & 15.84 \\
$U$ (1.5 M) & 10.55 & 14.51 & \textbf{17.84} & 20.68 & 11.44 & 15.41 \\
$U^*$ (3 M)& 11.48 & 15.46 & 17.60 & \textbf{21.03} & 12.29 & 16.08 \\
$U\rightarrow D$ & 11.96 & 15.50 & 15.08 & 18.37 & 12.49 & 16.01 \\
$D\rightarrow U$ & 14.09 & 16.97 & 11.13 & 17.59 & 13.42 & 16.95 \\
\midrule
$H$ (Ours) & 13.81 & 16.78  & 15.10 & 19.09 & \textbf{14.02} & \textbf{17.08} \\
\bottomrule
\end{tabular}
\end{center}
\label{table:main}
\end{table}

\subsection{Comparison with Other Methods}
Using the test dataset, we compared our results to recent studies of speech enhancement field. With the same training and test dataset split, our model showed the best performance quantitatively and qualitatively among the others under various measures \cite{Hu2008EvaluationOO} (Table \ref{table:best}). 

For the qualitative results, please refer to the web demo page (\url{https://mdphdnet.github.io}), where we have uploaded several denoised examples using the models introduced in the table (except MMSE-GAN whose code is unavailable).

\begin{table}[htb]
\caption{Comparison with other methods. Note that, the indicators that are not reported in the original paper are marked as dash (-). The predicted rating of speech distortion (CSIG), background distortion (CBAK) and overall quality (COVL) are reported (from 1 to 5, higher is better). PESQ (from -0.5 to 4.5, higher is better) stands for perceptual evaluation of speech quality and SSNR (higher is better) is segmental SNR. The best result for each measure is given in bold style.
} 
\begin{center}
\begin{tabular}{lccccc}
\toprule
\multicolumn{1}{c}{}  &\multicolumn{1}{c}{\bf CSIG}  &\multicolumn{1}{c}{\bf CBAK}  &\multicolumn{1}{c}{\bf COVL}  &\multicolumn{1}{c}{\bf PESQ} &\multicolumn{1}{c}{\bf SSNR}
\\ \midrule
Wiener \cite{scalart1996speech}  & 3.23 & 2.68 & 2.67 & 2.22 & 5.07 \\
SEGAN \cite{Pascual2017segan}    & 3.48 & 2.94 & 2.80 & 2.16 & 7.73\\
Wavenet \cite{rethage2017wavenet}  & 3.62 & 3.23 & 2.98 & - & -\\
MMSE-GAN  \cite{soni2018time}  & 3.80 & 3.12 & 3.14 & 2.53 & -\\
TasNet (3M) & 3.80 & 3.29 & 3.18	& 2.57 & 9.65 \\ 
U-Net (3M) & 3.65 & 3.21 & 3.05	& 2.48 & 9.34 \\ 
MDPhD (3 M + 3 M) & \textbf{3.85} & \textbf{3.39} & \textbf{3.27} & \textbf{2.70} & \textbf{10.22}\\
\bottomrule
\end{tabular}
\end{center}
\label{table:best}
\end{table}

\section{Conclusion}

We demonstrated that the conventional speech enhancement models have limitations due to using specific representations. Based on this observation, we proposed a hybrid approach that exploits multi-domain features for speech enhancement, dubbed multi-domain processing via hybrid denoising networks (MDPhD). With respect to five metrics, MDPhD achieved the best performance compared to the other concurrent models. Because MDPhD is a general framework, future work may include developing a more elegant way of hybridizing and extending this framework to other signal processing tasks, such as music and speech source separation.

\bibliographystyle{plain}
\bibliography{ref}

\newpage
\clearpage

\appendix
\section{More Experimental Results}

To show that the complementary nature of the time domain and T-F domain networks does not come from a specific choice of the objective functions, we trained each network module with various objective functions. Table \ref{table:ablation_lossmask} summarizes the results. We found that the performance was not significantly different.

\begin{table}[h]
\caption{SNR evaluation of models with various objective functions. $D$ and $U$ denote the TasNet (reduced) using one-dimensional dilated convolution and U-Net, respectively. The type of objective functions are noted next to the model name. $\ell_1$ represents our baseline objective function. $\ell_2$ represents an objective function that substitutes the $\ell_1$ term of equation (1) with $\ell_2$. SNR indicates an objective function that directly optimizes the SNR. SPEC represents the $\ell_2$ distance between a clean speech spectrogram and the estimated spectrogram.}
\renewcommand{\tabcolsep}{0.55mm}
\begin{center}
\begin{tabular}{l|ccc|ccc}
\toprule
\multicolumn{1}{c}{} &\multicolumn{3}{|c}{\bf babble noise} &\multicolumn{3}{|c}{\bf high frequency band}
\\ \midrule
\multicolumn{1}{c}{} &\multicolumn{1}{|c}{\ SNR 5} &\multicolumn{1}{c}{\ SNR 10} &\multicolumn{1}{c}{\ SNR 15}&\multicolumn{1}{|c}{\ SNR 5} &\multicolumn{1}{c}{\ SNR 10} &\multicolumn{1}{c}{\ SNR 15 }
\\ \midrule
$D$-$\ell_1$& 13.69 & 16.83 & 19.57 & 4.96 & 11.31 & 16.54\\
$D$-$\ell_2$ & 13.53 & 16.57 & 19.26 & 6.67 & 12.82 & 16.94\\
$D$-SNR & 13.45 & 16.71 & 19.51 & 4.48 & 10.90 & 16.38\\
$U$-$\ell_1$ & 10.55 & 14.51 & 18.11 & 17.87 & 20.68 & 21.92 \\
$U$-$\ell_2$ & 10.54 & 14.48 & 17.97 & 17.89 & 20.65 & 22.32\\
$U$-SPEC & 10.47 & 14.38 & 18.01 & 19.73 & 21.47 & 22.27\\
\bottomrule
\end{tabular}
\end{center}
\label{table:ablation_lossmask}
\end{table}

\section{Model Architecture}
In this section, we present the detailed configuration of the models we used. In the following figures, each block consists of a convolutional operation, normalization and an activation function. Note that, normalization is not used at the first and the last layer of each model. The operation $\odot$ means element-wise multiplication and the preceding layer of this operation uses sigmoid as an activation function.

\begin{figure}[h]
\centering
{\includegraphics[width=0.70\linewidth]{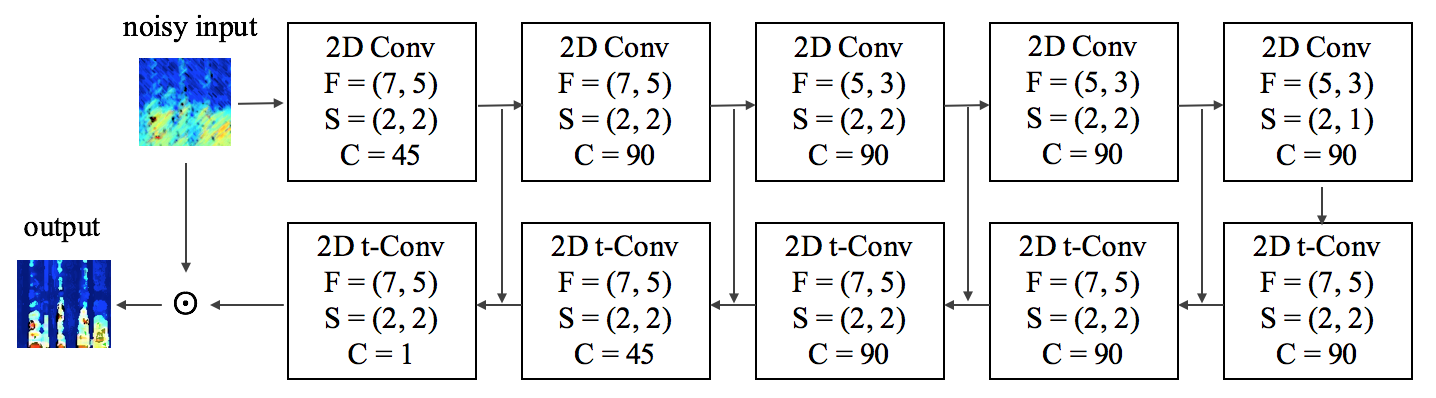}} \\
\caption{U-Net (1.5M) architecture. 2D Conv means a two-dimensional convolution block consisting of a two-dimensional convolution operation with filter size F (height, width), stride size S (height, width) and output channel size C followed by batch renormalization and leaky-RELU activation function. 2D t-Conv means a two-dimensional transposed convolution block. Our baseline models used in experiments process the log-magnitude of the input spectrogram.}
\label{fig:unet1.5M}
\end{figure}

\begin{figure}[h]
\centering
{\includegraphics[width=0.99\linewidth]{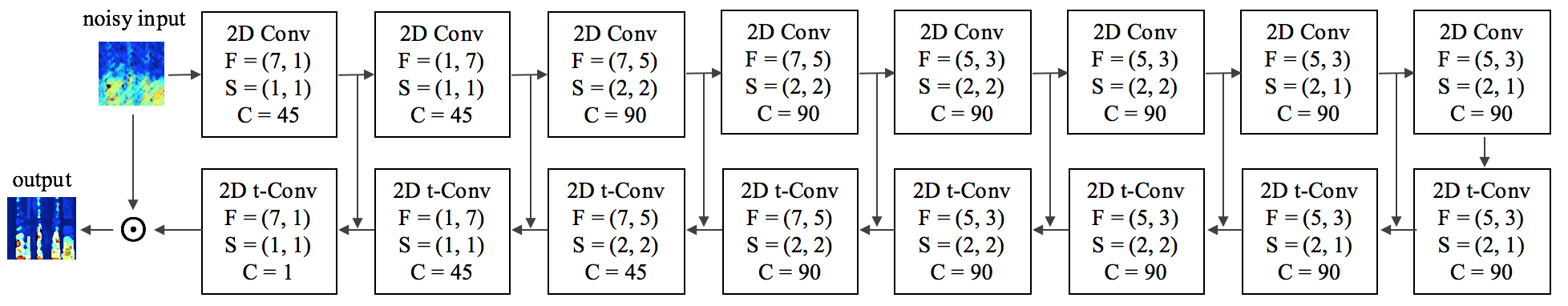}} \\
\caption{U-Net (3M) architecture.}
\label{fig:unet3M}
\end{figure}

\begin{figure}[htbp]
\centering
{\includegraphics[width=0.99\linewidth]{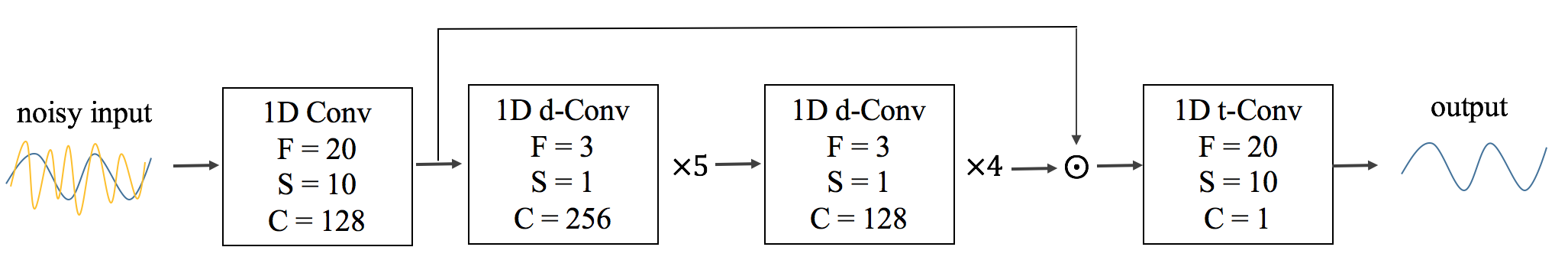}} \\
\caption{TasNet (1.5M) architecture. 1D Conv means a one-dimensional convolution block and 1D d-Conv stands for a one-dimensional dilated convolution block. The dilation rate of each dilated convolution block is doubled as it goes forward. The convolution operation of the dilation convolution block follows the non-causal method, which takes the value of both ahead and back of the current time step. 1D t-Conv means a one-dimensional transposed convolution block.}
\label{fig:tasnet1.5M}
\end{figure}

\begin{figure}[htbp]
\centering
{\includegraphics[width=0.99\linewidth]{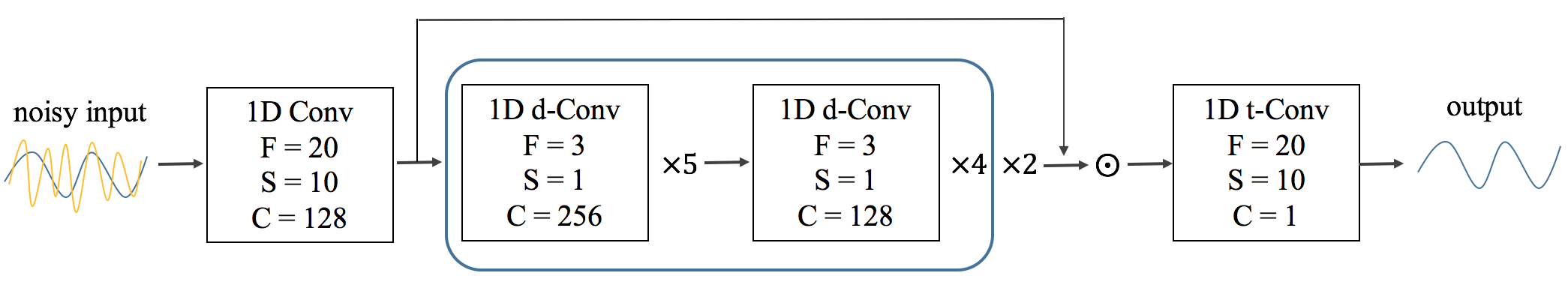}} \\
\caption{TasNet (3M) architecture.}
\label{fig:tasnet3M}
\end{figure}

\small
\end{document}